\documentclass[11pt,twoside]{article}

\usepackage{asp2006}
\usepackage{fancyhdr,graphicx,caption,rotating,multirow,lscape}
\DeclareGraphicsExtensions{.eps,.eps.gz,.ps,.jpg,.tiff}

\begin{document}

\title{Transit and secondary eclipse photometry\\ in the near-infrared}  



\author{Ignas Snellen}
\affil{Leiden Observatory, Leiden University, Postbus 9513, 2300 RA, Leiden, The Netherlands}


\begin{abstract} 
Near-infrared photometry of transiting extrasolar planets can be of great scientific value. It is however not straightforward to reach the necessary  millimagnitude precision. Here we report on our attempts to observe transits and secondary eclipses of several extrasolar planets at 2.2 $\mu$m. Best results have been obtained on OGLE-TR-113b using the SOFI near-infrared camera on ESO's New Technology Telescope. Its $K$-band transit shows a remarkably flat bottom indicating low stellar limb darkening. Secondary eclipse photometry has resulted in a formal 3$\sigma$ detection, but residual systematic effects make this detection rather uncertain. 
\end{abstract}

\section{Near infrared observations of transiting planets}   

Near-infrared photometry of transiting extrasolar planets can be of great scientific value. First of all, since the emergence of wide field near-infrared cameras, 
transit surveys in J-band targeting M dwarfs have become feasible 
(see Hodgkin et al., this volume). The areas of their stellar disk 
are up to two orders of magnitude smaller than that of the sun,
enabling detections of transits down to Earth-size planets. 
Furthermore, due to their low luminosities, the habitable zones around M dwarfs
are significantly closer to the host star, increasing the transit probability 
for life-bearing planets. 

\begin{figure}[!ht]
\centering
\includegraphics[angle=0,width=13cm]{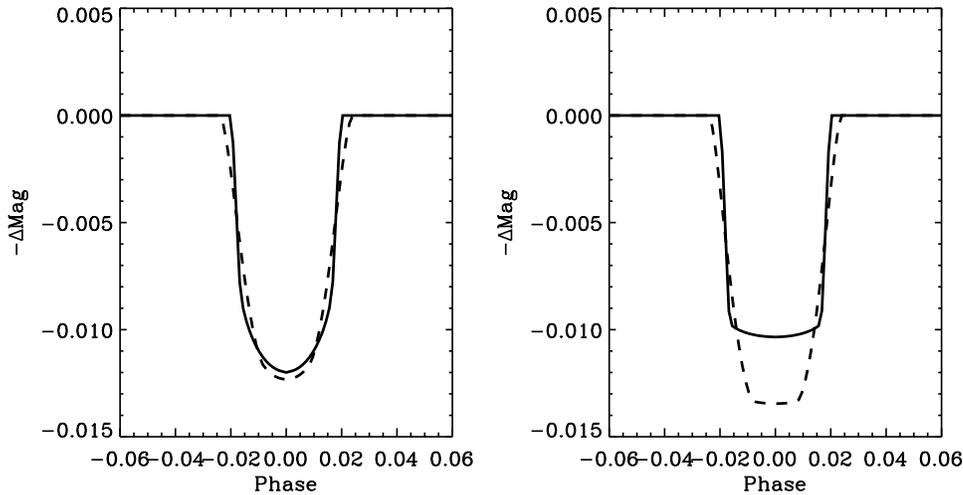}
\caption{ \label{snellen_fig1} Theoretical light curves of two eclipsing systems with 3 day periods. The dashed line is that from a mid M-dwarf eclipsing an A star with a high impact parameter, and the solid line that of a Jupiter size planet transiting a solar type star with a low impact parameter. The panel to the left is in V-band, where both light curves are hardly distinguishable. The panel to the right is in K-band, where the light curve of the eclipsing binary system is significantly deeper.}
\end{figure}

In addition, follow-up near-infrared transit photometry of known extrasolar 
planets is also valuable. The main difference from the optical is that the stellar limb darkening is significantly reduced, in 
particular at K-band. This results in a stronger contrast between the 
ingress/egress periods and the main part of the transit, allowing a 
better estimate of the transit impact parameter and the planet/star size ratio.
This can also help to distinguish false interlopers among planet candidates 
in transit surveys. Figure \ref{snellen_fig1} shows the model $V$ and $K$ band light curves of two systems, one of a mid M-dwarf transiting an A star at low impact, and the other of a Jupiter size planet transiting a G dwarf at high impact. While in the optical the two light curves are virtually identical, in $K$ band they are easily distinguishable. The low impact transit of the M dwarf is 
significantly deeper in $K$ due to the lower limb darkening, making the occulted surface area of the primary relatively brighter.

Although it puts the strongest demands on photometric accuracy, arguably most interesting are observations of the secondary eclipse of transiting planets. 
Recent detections of the secondary eclipses of TrES-1b, HD209458b, HD149733b, 
and others  using the Spitzer space 
telescope at mid-infrared wavelengths, constitute the first detections of 
direct thermal light from extrasolar planets (Charbonneau et al. 2005; Deming et al. 2005; Deming et al. 2006; J. Harrington et al., this volume).
These data show that hot Jupiters have surface temperature of 1000-1500 K, 
confirming the predicted heating by stellar irradiation (Seager and Sasselov
 1998). Detailed interpretation of the infrared measurements is however 
difficult, since e.g. it is not evident to what extent they are influenced by water
 vapour opacity. If a hot Jupiter would radiate as a black body with a 1000-1500 K surface temperature, one would expect a $K$-band secondary eclipse with a depth in the order of maybe a few times 100 $\mu$mag, probably beyond the reach of current instrumentation. However, the near-infrared spectrum of these planets are expected to be dominated by broad absorption features due to 
CO and H$_2$O, suppressing the light in these bands, but enhancing the spectrum
in between the absorption. This results in strong spectral peaks, making them 
significantly brighter than expected from their black body temperature, 
in particular at 2.2$\mu$m(eg. Seager et al. 2005; Fortney, this volume). 
Secondary eclipse measurements at $K$ band should therefore not only be 
possible,  but are also important 
 to test detailed atmosphere models, and to interpret the planet's spectrum in 
terms of atmosphere temperature and Bond albedo.

\section{Technical challenges and first results}

In the remainder of this paper we concentrate on high precision photometry 
at 2.2 $\mu$m. While $K$-band secondary eclipse signals of hot or very hot 
Jupiters are not expected to exceed a few milli-magnitude, exoplanet transit 
signals can have depths of the order of a percent or more. In both cases 
however one would require 10$^{-3}$ precisions per 10 minutes or better,
to either obtain an indisputable detection of a secondary eclipse, or 
to precisely measure a transit profile. 

Considering only the photon noise coming from the star/planet system,
this should not be a problem. A 4m telescope receives about 1$\times$10$^7$
photons per second in K-band from HD209458a/b, while it still receives  2$\times$10$^{6}$ photons per minute from a much fainter system like OGLE-TR-113. 
Unfortunately, the photometric accuracy at $K$-band is dominated by 
 time dependent instrumental effects, such as possible 
systematic errors in illumination and flat fielding, intra-pixel sensitivities,
dark currents, PSF variability, and non-linearity effects. 
This, in addition to possible rapid fluctuations in atmospheric 
conditions, makes it rather challenging to obtain 10$^{-3}$ noise levels.

First attempts by Snellen (2005) to reach milli-magnitude photometric 
precision at 2.2 $\mu$m involved observations of the secondary eclipse of HD209458b 
using the UK InfraRed Telescope (UKIRT). The telescope was defocused to
avoided saturation of the K=6.3 star, and to diminished the systematic 
uncertainties in the photometry. The drawback of HD209458 as a target  is that 
no nearby, similarly bright comparison stars are available to perform 
relative photometry. Therefore observing cycles between the target
and reference stars were needed, limiting the accuracy due to atmospheric 
absorption and point-spread-function/seeing variations at the time-scale of 
the cycle time. Although the secondary eclipse was not detected, it was 
however shown that a photometric precision of 0.1\% is possible (Snellen 2005).

A second attempt to detect the secondary eclipse of a transiting extrasolar planet at $K$-band
involved DDT ISAAC/VLT observations of OGLE-TR-56. The rationale of using this much fainter
target was that photon noise would not be a problem anyway, and that in this way many comparison
stars of similar magnitude would be available instantaneously on the detector array. Unfortunately these
observations were not a success. The OGLE-TR-56 field is overcrowded, which in combination with 
a rapidly changing point spread function in both time and spatial position on the array, made 
accurate photometry impossible. In addition, it was not allowed to move the VLT more often than
once a minute, making the execution of a fast jitter pattern not possible. The most accurate relative photometry performed using this data set resulted in even bright stars on the array to apparently vary by 1$-$4\% in flux density on a 1$-$2 hour time scale. 

\section{Lessons learned: SOFI/NTT observations of OGLE-TR-113b}

\begin{figure}[!ht]
\centering
\includegraphics[angle=0,width=13cm]{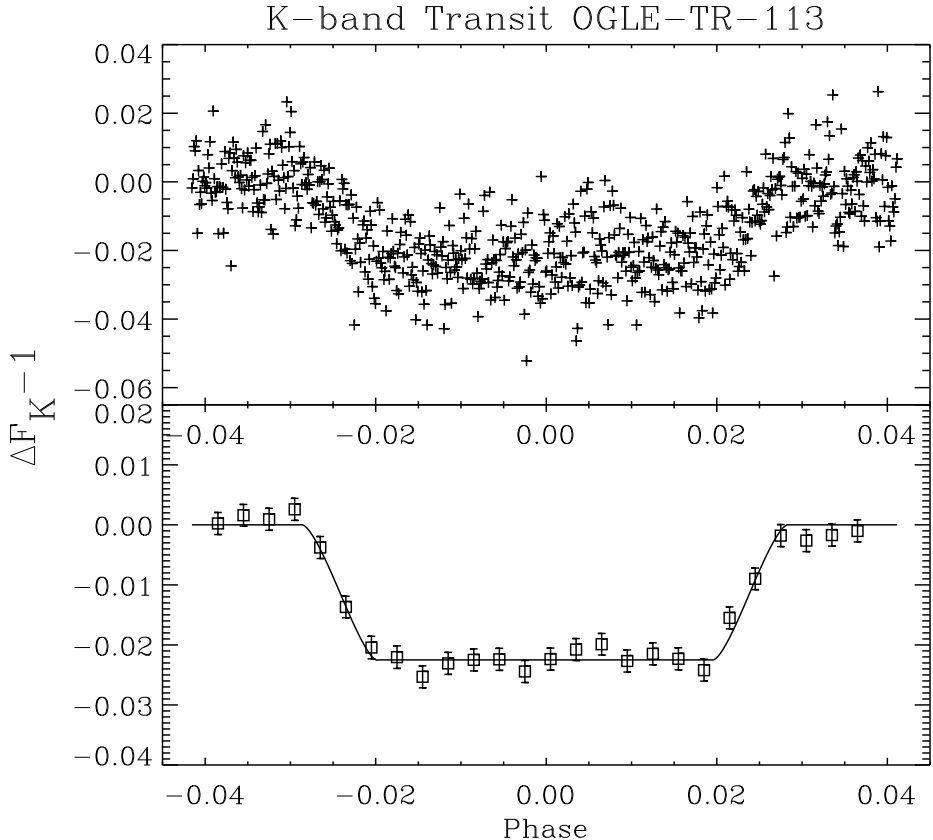}
\caption{ \label{snellen_fig2} From Snellen \& Covino 2007. K-band transit light curve of OGLE-TR-113. Note its remarkably flat bottom implying low limb darkening at this wavelength.  }
\end{figure}
\begin{figure}[!ht]
\centering
\includegraphics[angle=0,width=12cm]{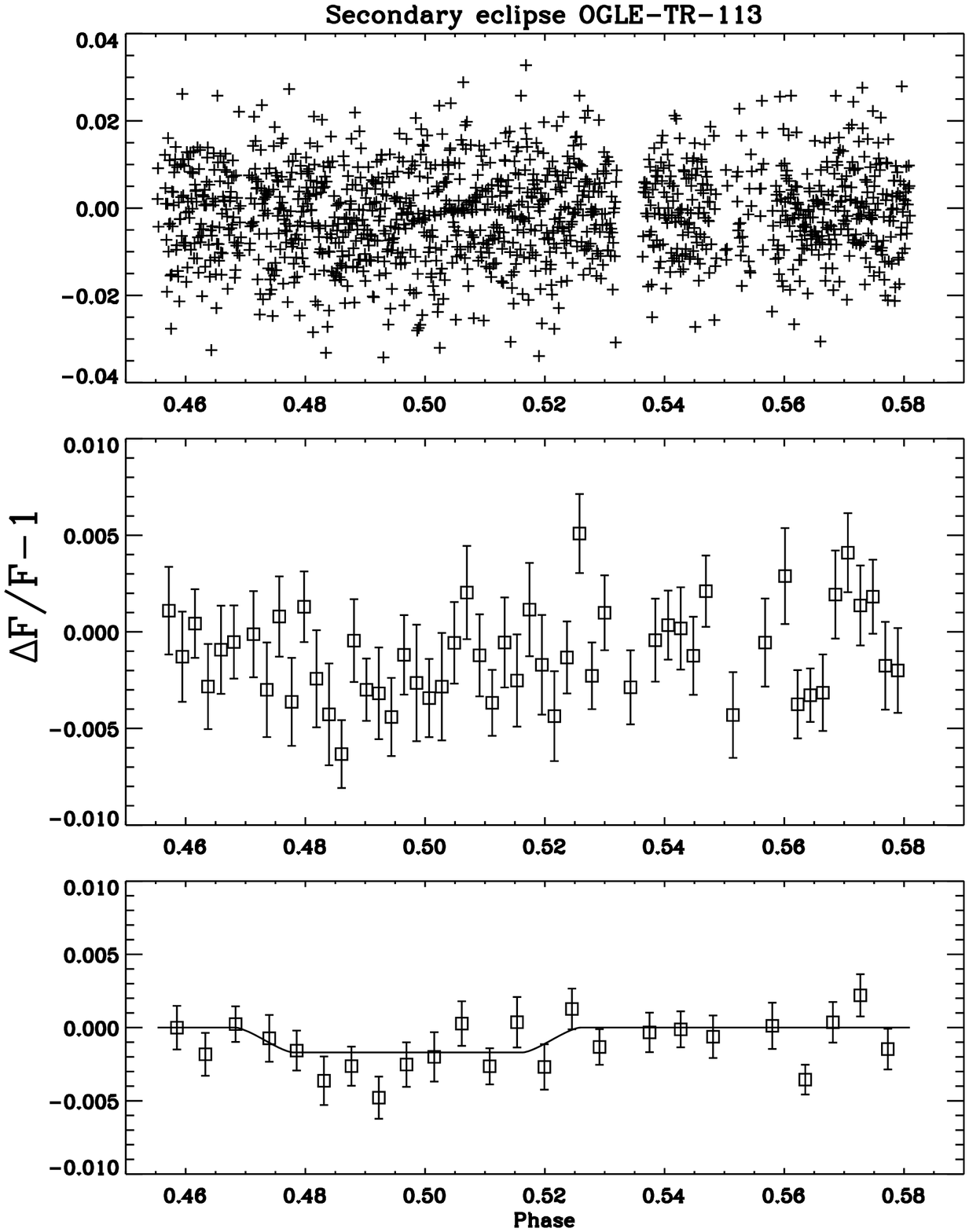}
\caption{ \label{snellen_fig3} From Snellen \& Covino 2007. The secondary eclipse light curve of OGLE-TR-113 in K-band. The eclipse is statistically detected at 2.8$\sigma$, however residual correlated noise  makes this detection rather tentative.}
\end{figure}

From our previous experiences we became uncertain whether, due to systematic effects inherent to 
K-band imaging,  it is possible to obtain a photometric accuracy better than $\sim$1\% per 
frame at all. We therefore chose for our next observing run a different strategy. These observations involved 
$K$-band transit and secondary eclipse photometry on OGLE-TR-113b, using SOFI on ESO's NTT,
which are described in detail in Snellen \& Covino (2007). For these observations we assumed that each individual frame would result in a photometric measurement with a precision of $\sim$1\%, but that by taking as many exposures within a given time as possible, each with random jitter position offsets, the noise would decrease to 1\%/$\sqrt{N}$. The exposure times are chosen to be as short as possible, but long enough to keep the photon noise from the target and background well below 1\%. 
We show below that this observing strategy worked well.

\subsection{Results: transit photometry}

As described in detail by Snellen \& Convino (2007), 665 $K$-band images of OGLE-TR-113 were taken in 3 hours, centered on the transit of March 17, 2006, each with an exposure time of 2$\times$5 seconds. The resulting light curve  (of individual data points and in 6 minute bins) is shown in figure \ref{snellen_fig2}. The 665 independent flux measurements have a dispersion of 0.9\% around the best fitting model, while the 6 minute binned data
show a scatter of 0.16\%. The transit is remarkably flat bottomed, indicating a low stellar limb darkening coefficient in $K$-band.

\subsection{Results: secondary eclipse photometry}

Observations on March 15, 2006, resulted in 1337 images in 4.3 hours, all taken with a 5 second exposure time. The resulting light curve is shown in figure \ref{snellen_fig3}.
The unbinned data shows a dispersion of 1.1\%, while the 5 and 10 minute binned data show a 
scatter of 0.24\% and 0.17\% respectively. The secondary eclipse of OGLE-TR-113 is statistically detected at a level of 2.8$\sigma$ at 0.17\%$\pm$0.05\%. However, as one can see best from the 5 minute binned light curve, some low level systematic effects still seem to be present on time scales of 5-10 minutes. The combination of the low significance of the eclipse, and the presence of this correlated noise, makes this detection rather tentative (Snellen \& Covino 2007). 

\section{Conclusions}

Near-infrared observations of transiting extrasolar planets can be of great scientific value, if 
a photometric precision at milli-magnitude level can be achieved. While our experiences show that 
it is difficult to reach an accuracy better than 1\% per exposure, a significantly better overall precision can be reached by using a combination of short exposures and a random jitter pattern.

\subsection*{Acknowledgments}

Part of the observations in this paper were collected at the European Southern Observatory, La Silla, Chile, using SOFI at the 3.5m NTT, within the observing program 076.C-0674(A), and at Paranal using ISAAC at the VLT within the observing program 275.C-5018 (DDT). the United Kingdom Infrared Telescope (UKIRT) is operated by the Joint Astronomy Centre on behalf of the U.K. Particle Physics and Astronomy Research Council.





\end{document}